\documentclass[a4,12pt]{article}

\newcommand{\lam}{\mbox{$\rm \Lambda$}}% Lambda
\usepackage{graphicx}
\usepackage{here}

\begin{document}

\title{ Polarized SIDIS: comment on  purity method for extraction 
of polarized quark distributions \footnote{The main idea of this note
has been presented in~\cite{ak}.}} 

\author{Aram~Kotzinian \\[1.cm]
Yerevan Physics Institute, 375036 Yerevan, Armenia\\
and JINR, 141980 Dubna, Russia\\
email: aram.kotzinian@cern.ch}

\maketitle

\begin{abstract}
The role of hadronization mechanism in polarization phenomena in 
semi inclusive deep inelastic scattering (SIDIS) and a purity method 
for extraction of polarized distribution functions are discussed. 
According to 
the  Monte Carlo (MC) event generator, on which this method is based,
hadrons can be produced  via quark (diquark) fragmentation or light
cluster decays. In contrast, the purity method assumes that only quark
fragmentation gives contribution to hadron production in the current
fragmentation region. The ignorance of contributions from diquark
fragmentation and cluster decays to asymmetry can be source of
incorrect values of polarized quark distributions 
extracted by the purity method.
\end{abstract}

%\newpage

The investigation of deep inelastic scattering (DIS) of
leptons is an excellent source to study the internal structure of
nucleon.  We know with good precision the unpolarized parton
distribution functions. Many efforts has been done to
investigate also the polarized parton content of nucleon. In this
case, for example, to separate the individual polarized  parton
distributions in the LO we still need more data ($g_1$ measurement at
very high $Q^2$, or asymmetry measurements in neutrino DIS on polarized
target) or some additional input like information from hyperon
$\beta$-decay.   
        Experimental progress in recent years allows to
investigate SIDIS. There is hope that, for
example, the measurement of different hadron production asymmetries on
proton and neutron targets will allow a further flavor separation of
polarized quark distributions. 

Recently the new preliminary data from HERMES on quark flavor separation 
has been presented~\cite{pur}. The LO analysis of semi inclusive DIS
has been done by using the purity method and suggests
that ``the strange sea appears to be positively
polarized'' in contrast to generally accepted 
negatively polarized strange sea scenario at LO (see discussion 
in~\cite{ls}). An ``explanation'' of HERMES result of a
positively polarized strange sea is proposed in~\cite{bass}. 
The idea is in transverse momentum dependence of polarized
photon-gluon fusion process $\gamma^* g \rightarrow q \bar q$. 
In my opinion this explanation can not be accepted by two reasons. 
First, the quantity
$g_1^{\gamma^* g \rightarrow q \bar q}$ for $q=s$ can not be
interpreted as a strange see contribution to nucleon spin. 
Second, HERMES made a LO analysis of the measured asymmetries.

It is evident that theoretical description of SIDIS is much more
complicated than DIS due to our poor knowledge of nonperturbative
hadronization mechanism.
Traditionally one distinguishes two regions for hadron
production: the  current fragmentation region, $x_F>0$ and 
the  target fragmentation region, $x_F<0$.
The  common assumption is that when selecting hadrons in the current 
fragmentation region and imposing a cut $z>0.2$ 
we are dealing with the quark fragmentation. 

To make flavor decomposition into polarized quark distributions 
the purity method 
has been used in the HERMES analysis~\cite{pur}. 
In the LO approximation the virtual photon asymmetry for hadron $h$
production is given by

\begin{equation}
\label{eq:a1}
A_1^h \simeq \frac{\sum_q e_q^2 \Delta q(x) 
\int_{z_{min}}^1 dz D_q^h(z)}
{\sum_q e_q^2 q(x) \int_{z_{min}}^1 dz D_q^h(z)}.
\end{equation}
This equation can be rewritten in the form

\begin{equation}
\label{eq:a1pur}
A_1^h \simeq  \sum_q P_q^h(x)\frac{\Delta q(x)}{q(x)},
\end{equation}
where the purity, $P_q^h(x)$, is defined as

\begin{equation}
\label{eq:pur}
P_q^h(x)=\frac{e_q^2 q(x) \int_{z_{min}}^1 dz D_q^h(z)} {\sum_{q'}
  e_{q'}^2 q'(x) \int_{z_{min}}^1 dz D_{q'}^h(z)}, 
\end{equation}
and calculated using an unpolarized MC event
generator {\tt LEPTO}~\cite{lepto}. 
Then, using measured asymmetries for different hadrons one can find
$\Delta q(x)$ by solving Eq.~\ref{eq:pur}. The main assumption of this
method is that all hadrons in the current fragmentation region with $z>0.2$ 
are produced from the quark fragmentation and
there is no additional terms in both the numerator and the denominator of
Eqs.~\ref{eq:a1} and~\ref{eq:a1pur}. 

However, this assumption fails for moderate energies of current fixed
target experiments. In {\tt LEPTO} generator 
hadronization is performed by {\tt JETSET}~\cite{JETSET} program which
is based on Lund fragmentation model. 
In this program there is a pointer which
shows the origin of produced hadrons. They can be originating from the
quark or target remnant diquark fragmentation or from small mass
cluster isotropic decay. 

\begin{figure}[hbt]
\begin{center}
\vspace {-0.5cm}
 \includegraphics[width=95mm]{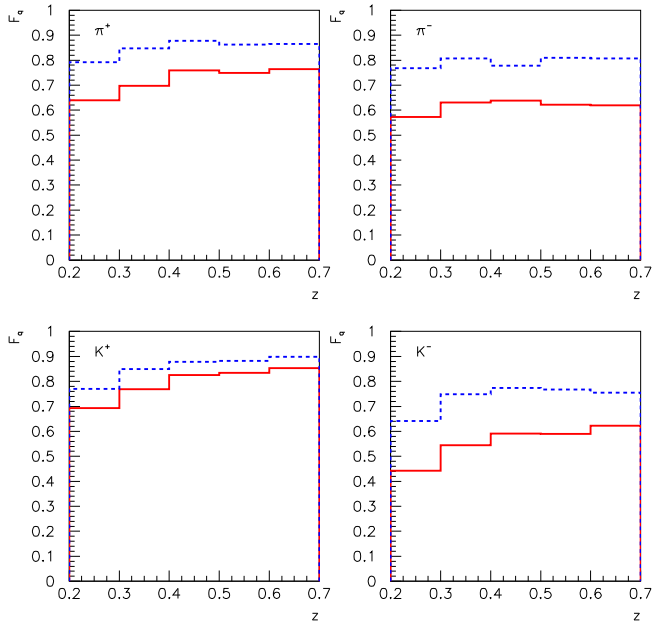}
\vspace {-0.7cm}
\caption{\label{fig:hzxf}Fraction of hadrons originating from quark
  fragmentation. Full line -- hadrons are selected with
  cuts $x_F>0$ and $z>0.2$, dashed line -- additional cut $W>6GeV$ is
  also imposed.}
\end{center}
\vspace {-0.5cm}
\end{figure}

Let us consider a sample of generated with 
{\tt LEPTO} events for HERMES energy and select hadrons in the
current fragmentation region. The fraction of events with hadrons 
produced via quark fragmentation,
$$
{\tt F_q=\frac{N_{hadron}({from \;quark \;fragmenation})}{N_{hadron}(tot)}},
$$
is presented in Fig.~\ref{fig:hzxf} for different hadrons as a
function of $z$. As one can see this fraction has a weak dependence on
$z$ and is less than {\it one} even at large values of $z$. With
additional cut on the final hadron system mass, $W>6GeV$ this
fraction becomes higher, but it is still about 20\% less than {\it one}. 
From this
consideration I conclude that even if data on asymmetries show little
dependence on $z$ this doesn't mean that we deal with the independent
quark fragmentation.

Thus, the assumption that hadrons in the current fragmentation region 
are produced only via quark fragmentation is not
valid in the Lund model and purities obtained with {\tt LEPTO}
generator include contributions from the target remnant
and cluster fragmentation. 
Taking into account these contributions the SIDIS cross section 
in general can be expressed as 
\begin{equation}
\label{eq:dsig}
d\sigma_{p}^h \propto {\sum_q [q(x)D_q^h(z)+M_q^{h/p}(x,z)]}.
\end{equation}
The last term has to be considered as contribution from fracture
functions introduced in~\cite{tv}. The same holds also for polarized
case. Then, Eq.~\ref{eq:a1} for virtual photon asymmetry is
modified: 
\begin{equation}
\label{eq:a1frac}
A_{1p}^h=\frac{\sum_q [\Delta q(x)D_q^h(z)+\Delta M_q^{h/p}(x,z)]}
{\sum_q [q(x)D_q^h(z)+M_q^{h/p}(x,z)]}.
\end{equation}
The additional contributions from diquark fragmentation and other
sources arise in numerator and denominator. 

To investigate the stability of purity method the following MC
exercise has been done.  Using {\tt PEPSI} polarized MC 
event generator~\cite{pepsi} the sample of $10^8$ 
SIDIS events has been generated  at HERMES energy for each relative 
polarization state of beam and target. 
The GRSV2000 LO (standard scenario)~\cite{grsv} polarized 
and corresponding GRV98 LO~\cite{grv} unpolarized distribution
functions have been used. The events with  
$\pi^+, \pi^-, K^+, K^-, h^+$ and $h^-$ (here $h$ stands for
hadrons without type identification) in the current fragmentation 
region with $x_F>0$ and $z>0.2$ have been selected. 
It is known that unpolarized data are well
described by MC event generator (denominator in Eq.~\ref{eq:a1frac}).
In contrast, nothing is known about polarized fracture functions
$\Delta M_q^{h/p}(x,z)$ 
(numerator in Eq.~\ref{eq:a1frac}). As an example, the asymmetry 
$$A_h^{obs}=\frac{N_h^{\uparrow\downarrow}-N_h^{\uparrow\uparrow}}
{N_h^{\uparrow\downarrow}+N_h^{\uparrow\uparrow}}$$
is calculated under two different assumptions for numerator:
\begin{itemize}
\item [--] {\bf Model 1:} all selected generated events gives
  contribution into numerator. 
In this case $\Delta M_q^{h/p}(x,z) \propto \Delta q(x)$.
 
\item [--] {\bf Model 2:} only hadrons from the quark fragmentation gives
  contribution into numerator ($\Delta M_q^{h/p}(x,z)=0$).
\end{itemize}

\begin{figure}[htb]
\begin{center}
\vspace {-2.5cm}
 \includegraphics[width=120mm]{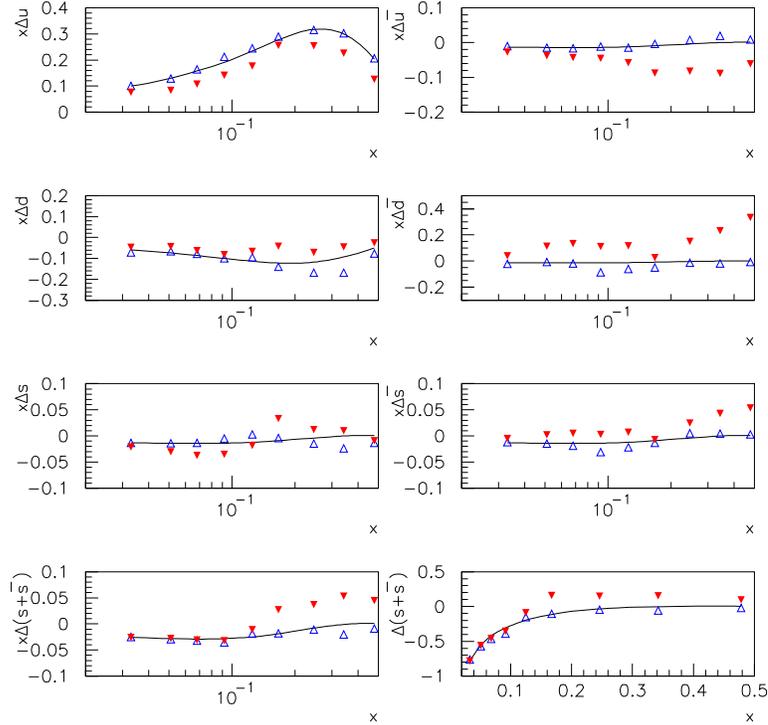}
\vspace {-3.cm}
\caption{\label{fig:xdq}
Reconstructed by purity method polarized quark distributions as a 
function of $x_{Bj}$: 
empty triangles -- {\bf Model 1}, full triangles -- {\bf Model 2}, 
solid line -- the input distribution.}
\end{center}
\vspace {-0.2cm}
\end{figure}

Then, the purities are calculated using unpolarized MC sample and 
Eq.~\ref{eq:a1pur} is solved. Reconstructed polarized distributions are
presented in Fig.~\ref{fig:xdq}. As one can see the two models
give very different results. Of course, the  second model has to be
considered as an extreme example which only was chosen to
demonstrate instability of purity method. In particular, with {\it
negative} input for polarized strange sea distribution the {\bf Model
2} leads to {\it positive} $\Delta s$. 

Recently the importance of target remnant effects in SIDIS has been
pointed out in many articles~\cite{gr}-\cite{ter}.
In~\cite{EKN} the model for \lam~hyperons longitudinal 
polarization in SIDIS has been developed. This model takes into
account that at moderate energies of fixed target experiments 
according to {\tt LEPTO} event generator \lam~hyperons are mainly 
originating from target remnant diquark fragmentation
even in the current fragmentation region.
Our model is able to describe all 
existing data on \lam~hyperons longitudinal polarization in SIDIS.
Within this model the NOMAD data~\cite{nomad} on \lam~hyperons
longitudinal polarization imply that the strange quarks inside the
nucleon has opposite polarization to that of struck valence quark.

In conclusion, I have demonstrated that due to additional sources 
of hadrons in the current fragmentation region the purity method 
can not be accepted as a precise tool for extraction of
polarized quark distribution. Within this method the ignorance of 
contributions from diquark fragmentation and cluster decays to 
asymmetry can bring to incorrect conclusions about light sea
polarization. At least, the uncertainties due to these effects 
have to be included into theoretical systematic errors in extracted
by purity method polarized quark distributions.

%\section*{Acknowledgements}


\begin{thebibliography}{99}

\bibitem{ak} A.~Kotzinian, to be published in the proceedings of \\
XVI ISHEPP, Dubna, 10--15 June, 2002;\\
X NATO Advanced Spin Physics Workshop, Yerevan 30 June--3 July, 2002;\\
Workshop on Future Physics @ Compass, CERN, 26--27 October, 2002.   

\bibitem{pur} H.E.~Jackson, Int. J. Mod. Phys. {\bfseries A17} 3551 (2002),
hep-ex/0208015;\\
A.~Miller, plenary talk at SPIN 2002,\\ 
www.c-ad.bnl.gov/SPIN2002/presentations/miller.pdf;\\
U.Stosslein, Acta Phys. Polonica {\bfseries B33} 2813 (2002);\\
M.~Beckmann, hep-ex/0210049.

\bibitem{ls} E.~Leader and D.B.~Stamenov, hep-ph/0211083.

\bibitem{bass} S.D.~Bass, hep-ph/0210214.

\bibitem{lepto} 
  G.~Ingelman, A.~Edin and J.~Rathsman,
  Comp. Phys. Commun. {\bfseries 101} 108 (1997).

\bibitem{JETSET} 
T.~Sj\"ostrand, {\it PYTHIA 5.7 and JETSET 7.4: physics
  and manual}, hep-ph/9508391; \\
  T.~Sj\"ostrand, Comp. Phys. Commun. {\bfseries 39} 347 (1986), 
{\bfseries 43} 367 (1987).

\bibitem{tv} L.~Trentadue and G.~Veneziano, Phys. Lett. {\bfseries
  B323} 201 (1994).

\bibitem{pepsi}
L.~Mankiewicz, A.~Schaefer and M.~Veltri, 
Comp. Phys. Comm. {\bfseries 71} 305 (1992),
http://www.thep.lu.se/~maul/pepsi.html.

\bibitem{grsv}M.~Gluck, E.~Reya, M.~Stratmann and W.~Vogelsang,
 Phys. Rev. {\bfseries D63} 094005 (2001). 

\bibitem{grv} M.~Gluck, E.~Reya and A.~Vogt, Eur. Phys. J. {\bfseries
C5} 461 (1998).

\bibitem{gr} M.~Gluck and E.~Reya hep-ph/0203063.

\bibitem{EKN} J.~Ellis, A.~Kotzinian, D.~Naumov, hep-ph/0204206.

\bibitem{br} S.J.~Brodsky, D.S.~Hwang and I.~Schmidt, Phys. Lett. 
{\bfseries B530} 99 (2002).

\bibitem{ter} O.V.~Teryaev, hep-ph/0211027.

\bibitem{nomad} P.~Astier {\it et al.}, [NOMAD Collaboration],
  Nucl. Phys. {\bfseries B588} 3 (2000).

\end{thebibliography}
\end{document}